\documentclass[journal,comsoc]{IEEEtran}
\usepackage[T1]{fontenc}
\usepackage{dsfont}
\ifCLASSINFOpdf
\else
\fi
\usepackage[linesnumbered, ruled]{algorithm2e}
\usepackage{amsmath}
\usepackage{multirow}
\usepackage{multicol}
\makeatletter
\newcommand{\removelatexerror}{\let\@latex@error\@gobble}
\def\ps@IEEEtitlepagestyle{%
	\def\@oddfoot{\mycopyrightnotice}%
	\def\@oddhead{\hbox{}\@IEEEheaderstyle\leftmark\hfil\thepage}\relax
	\def\@evenhead{\@IEEEheaderstyle\thepage\hfil\leftmark\hbox{}}\relax
	\def\@evenfoot{}%
}

\def\mycopyrightnotice{%
	\begin{minipage}{\textwidth}
		\centering \scriptsize
		This article has been accepted in IEEE Communications Letters Journal © 2021 IEEE. Personal use of this material is permitted. Permission from
		IEEE must be obtained for all other uses, in any current or future media, including reprinting/republishing this material for advertising or promotional purposes, creating new collective works, for resale or redistribution to servers or lists, or reuse of any copyrighted component of this work in other works. This work is freely available for survey and citation.
		
	\end{minipage}
}
\makeatother
\usepackage[cmintegrals]{newtxmath}
\usepackage{amsmath}
\usepackage{graphicx}

\begin{document}
%
\title{OSC-MC: Online Secure Communication Model for Cloud Environment }
%
%
%

\author{Deepika~Saxena and
        Ashutosh~Kumar~Singh,~\IEEEmembership{Senior~Member~IEEE}
\thanks{D. Saxena and A. K. Singh are with the Department of Computer Applications, National Institute of Technology, Kurukshetra, India. E-mail: 13deepikasaxena@gmail.com and ashutosh@nitkkr.ac.in }}

\markboth{IEEE COMMUNICATIONS LETTERS}%
{Shell \MakeLowercase{\textit{et al.}}: Bare Demo of IEEEtran.cls for Computer Society Journals}

\maketitle

\begin{abstract}
	A malicious cloud user may exploit outsourced data involved in online communication, co-residency, and hypervisor vulnerabilities to breach and hamper sensitive information, and inject malicious traffic based congestion, rendering services to other benign users. To address this critical and challenging problem, this letter proposes an Online Secure Communication Model for Cloud (OSC-MC) by identifying and terminating malicious VMs and inter-VM links prior to occurrence of security threats. The anomalous network traffic, bandwidth usage, and unauthorised inter-VM links are security breach indicators which guides secure cloud communication and resource allocation. The simulation and comparison of the proposed model with existing approaches reveal that it significantly improves authorised inter-communication links up to 34.5\% with reduction of network hogs, and power consumption by 66.46\% and 39.31\%, respectively.

\end{abstract}


%
\IEEEpeerreviewmaketitle

\section{Introduction}
%
%
%
%
\IEEEPARstart{T}{he} cloud platforms enabled with easy provisioning, rapid deployment, resource scalability, hardware consolidation etc., have  become critical avenues to many commercial, academic, and research organizations by providing them access to millions of customers at moderate cost \cite{opmlb}. However, the cloud communications suffer from high security breaches due to discrepancies and vulnerabilities associated with the networking devices, hypervisor and susceptibilities of side channels \cite{mlpam}, \cite{apo_risk}. A malicious user may inject network congestion, launch multiple VMs and exploit sequential or parallel VM placement to hamper real time communication and compromise benign user's VMs \cite{pssf}-\cite{allocation}. Therefore, securing online data execution and communication among different cloud users' applications, is a crucial and bottleneck problem. To the best of the authors' knowledge, there is no existing model which enables proactive protection against security threats via congestion,
co-location, network cascading effect, and vulnerability.
\par This letter proposes a novel \textbf{O}nline \textbf{S}ecure \textbf{C}ommunication \textbf{M}odel for \textbf{C}loud Environment (\textbf{OSC-MC}) to provide secure and network efficient data execution and transmission in cloud data centre networks. In this model, the inter-VM relations on a server are exclusively monitored, formulated, and analyzed as suspicious or non-suspicious by comparing with the authorized inter-communication links specified for a VM in a log maintained by a resource manager. OSC-MC investigates the inter-VM behaviour, identifies and suspends malicious VMs, and terminates all the malicious links before the completion of sufficient time required to accomplish successful attack. Furthermore, the incoming traffic is predicted online and distributed into clusters on the basis of bandwidth consumption of future applications to combat the network congestion based security attacks. The consecutive prediction and grouping of incoming traffic into clusters help in ($i$) determining the network hogs/congestion and mitigating them prior to occurrence, ($ii$) autoscaling of sufficient number of VMs and their scheduling on selected network-efficient servers with minimized bandwidth wastage. 
\section{OSC-MC}
\begin{figure}[!htbp]
	\centering
	\includegraphics[width=0.9\linewidth]{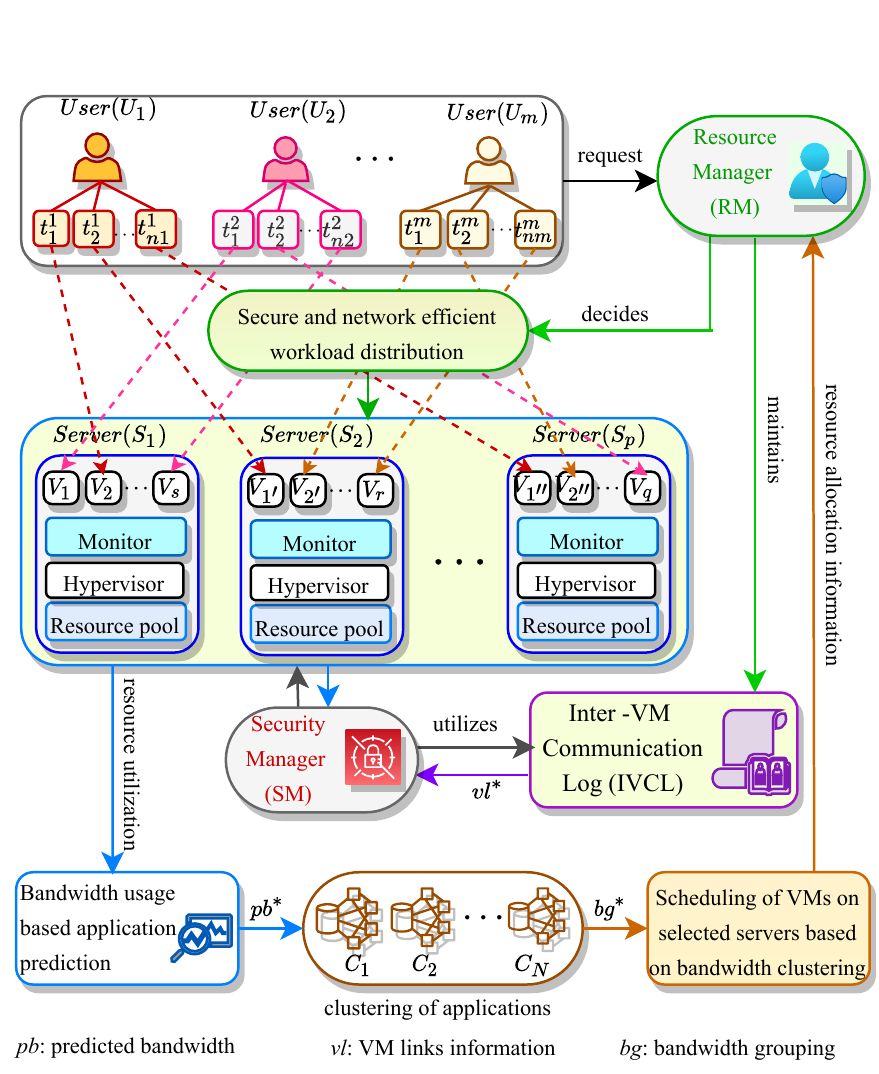}
	\caption{Online Secure Cloud Communication Model}
	\label{fig:figure-1-proposed-model}
\end{figure}

Consider $m$ users \{$U_1$, $U_2$, ..., $U_m$\}$\in \mathds{U} $ have requested execution of their application (i.e., Bag of Tasks \{$t_1$, $t_2$, ..., $t_n$\}) to \textit{Resource Manager} (RM) on a set of $q$ VMs \{$V_1$, $V_2$, ..., $V_q$\} $\in \mathds{V}$ hosted on $p$ servers \{$S_1$, $S_2$, ..., $S_p$\} $\in \mathds{S}$. For instance, VMs: \{$V_1$, $V_2$, ..., $V_s$\} are deployed on $S_1$; \{$V_{1'}$, $V_{2'}$, ..., $V_r$\} and \{$V_{1''}$, $V_{2''}$, ..., $V_q$\} on $S_2$ and $S_p$, respectively as shown in Fig. \ref{fig:figure-1-proposed-model}. 

RM distributes the users' tasks among the VMs purchased by them and allocates these VMs to the selected network-efficient (i.e., optimized bandwidth) servers. The periodic estimation of future resources (viz., CPU, memory, bandwidth etc.) utilization information of the VMs and predicted bandwidth usage based clustering of incoming traffic into groups assist the RM in deciding an efficient workload distribution while avoiding the future network congestion. Concurrently, RM generates and maintains \textit{Inter-VM Communication Log} ({IVCL}) which consists of the information of user specified authorized access links among the inter-dependent VMs based on the scheduling of their respective tasks. \textit{Security Manager} (SM) utilizes IVCL for investigation of possibility of any security threat by distinguishing unauthorized links among all the current inter-VM links on a server and terminating them beforehand. Each server consists of a \textit{hypervisor} and \textit{resource pool} to deploy different users' VMs and process tasks, respectively. An exclusive \textit{Monitor} layer is employed to surveil inter-communication links among VMs on the respective server. 
\section{Secure Inter-VM Communication  }
\begin{figure}[!htbp]
	\centering
	\includegraphics[width=0.9\linewidth]{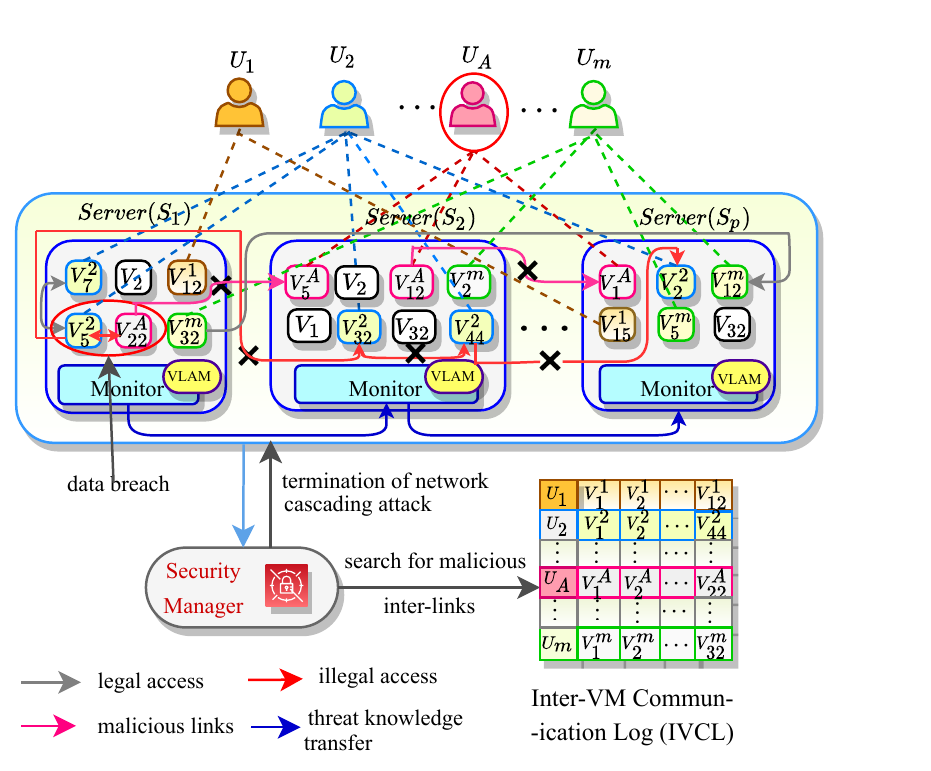}
	\caption{Secure inter-VM communication}
	\label{fig:figure-2-security-mechanism}
\end{figure}

Let the applications of users \{$U_1$, $U_2$, ... , $U_A$, ..., $U_m$\} are executed on different VMs \{${V}_i^k$: $i \in [1, q]$, $ k \in [1, m]$\} as illustrated in Fig. \ref{fig:figure-2-security-mechanism}. Each server deploys a \textit{Monitor} that maintains a \textit{VM Link Access Matrix} (VLAM) for the surveillance of the current inter-communication behaviour among the VMs hosted on it. Taking the assumptions: (\textit{i}) All of the VMs \{$V^A_1$, $V^A_2$, ..., $V^A_{22}$\} $\in V^{Mal}$ belonging to user $U_A$ are \textit{malicious VMs} and inter-links among them are \textit{malicious links}, (\textit{ii}) RM is unaware about \textit{malicious user} or \textit{attacker} ($U_A$), and (\textit{iii}) All the links established between a user VM and other known VM with access permission assigned by the respective user have \textit{authorized inter-communication links} ($\mathds{AL}$) which are recorded in IVCL. SM investigates the probability of threat by comparing the current information of VLAM with the information available in IVCL, periodically. Any mismatch between the observed inter-VM links and set of authorized inter-communication links is an indicator of security threat.   
\par Let $V_{i,z}$ and $V_{i^\ast,z^\ast}$ specify $i^{th}$ VM of $z^{th}$ benign user and ${i^\ast}^{th}$ VM of ${z^\ast}^{th}$ malicious user,  respectively and their mapping ($\omega$) on $k^{th}$ server is denoted as $\omega^z_{ik}$ and $\omega^{z^\ast}_{{i^\ast}k}$, respectively. The term $\omega_{ik}$=1, if $i^{th}$ VM is hosted on $k^{th}$ server, else, $\omega_{ik}$=0. The unauthorized inter-VM access $\Theta^{col}_{ik,z \rightarrow {i^\ast}k,z^\ast}$ from $V_{i,z}$ to $V_{i^\ast,z^\ast}$ co-located on $k^{th}$ server over time-interval \{$t_1$, $t_2$\} is stated in Eq \ref{threat1}; where $\uplus^V_{ik,z\rightarrow {i^\ast}k,z^\ast}$ specifies inter-VM relation between $V_{i,z}$ and $V_{i^\ast,z^\ast}$.  
\begin{equation} \label{threat1}
\int_{t_1}^{t_2} \Theta^{col}_{ik,z \rightarrow {i^\ast}k,z^\ast}dt = {\int_{t_1}^{t_2} (\omega_{ik}^z \times \omega_{{i^\ast}k}^{z^\ast} \times \uplus^V_{ik,z \rightarrow {i^\ast}k,z^\ast}})dt
\end{equation} 

The sets of authorized links ($\mathds{AL}_{legal}^{V_{i,z}}$) and current inter-VM links ($\mathds{L}^{V_{i,z}}$) specified for VM $V_{i,z}$ in access log (IVCL) and VLAM are stated in Eqs. (\ref{legallinks}) and (\ref{links}), respectively.
\begin{equation}
\{\mathds{AL}_1^{V_{i,z}}, \mathds{AL}_2^{V_{i,z}}, ..., \mathds{AL}_n^{V_{i,z}}\} \in \mathds{AL}_{legal}^{V_{i,z}}\label{legallinks}  \end{equation}
\begin{equation}
\{\mathds{L}_1^{V_{i,z}}, \mathds{L}_2^{V_{i,z}}, ..., \mathds{L}_n^{V_{i,z}}\} \in \mathds{L}_{}^{V_{i,z}}\label{links}  \end{equation}
 An inter-VM link between $V_{i,z}$ and $V_{i^\ast,z^\ast}$ placed on $k^{th}$ server is denoted as: $ \mathds{L}_{ik,z \rightarrow {i^\ast}k,z^\ast}$; where $\forall \{i, i^\ast\} \in [1,q]$, $i\ne i^\ast$, $k \in [1,p]$, $z \in [1,m]$ and the relationship ($\uplus^V_{ik,z\rightarrow {i^\ast}k,z^\ast}$) between them is revealed via comparison of VLAM and IVCL, is evaluated in Eq. (\ref{linkthreat}). The value of $\uplus_{ik,z\rightarrow {i^\ast}k,z^\ast}$ is $1$ for suspicious or unauthorized relation, otherwise, it is $0$.
\begin{equation}
\uplus^V_{ik,z\rightarrow {i^\ast}k,z^\ast} = \begin{cases}
$1$, & {If \quad {\mathds{L}_{ik,z \rightarrow {i^\ast}k,z^\ast}}} \nsubseteq \mathds{AL}_{legal}^{V_{i,z}}\\ 
$0$, & { {\textit{otherwise}} } 
\end{cases} \label{linkthreat} 
\end{equation} 
Eq. (\ref{cascading}) evaluates \textit{network cascading attack} ($ \Theta^{cas}_{({i}k,z\rightarrow {i^{\ast\ast}}k^\ast,z^\ast)}$) between VMs $V_{i,z}$ and $V_{i^{\ast\ast},z^\ast}$ deployed on $k^{th}$ and ${k^\ast}^{th}$ servers, respectively via $V_{i^\ast,z^\ast}$ on $k^{th}$ server, where $\forall \{i, i^\ast, i^{\ast\ast}\} \in [1,q]$, $k \in [1,p]$, $\uplus^V_{({{i}k},{z}\rightarrow {i^{\ast\ast}}k^\ast,z^\ast)}$ is defined using Eq. (\ref{relation})

\begin{equation} 
\resizebox{0.36\textwidth}{!}{$ 
\begin{aligned}
\label{cascading}
	\int_{t_1}^{t_2} \Theta^{cas}_{({i}k,z\rightarrow {i^{\ast\ast}}k^\ast,z^\ast)}dt = 	\int_{t_1}^{t_2} {\omega_{ik}^z \times \omega_{{i^\ast}k}^{z^\ast}\times \omega_{{i^{\ast\ast}}k}^{z} }\\ \times \uplus^V_{({{i}k},{z}\rightarrow {i^{\ast\ast}}k^\ast,z^\ast)} dt
	\end{aligned} $}
\end{equation}

\begin{equation}
\begin{aligned}
 \label{relation}
	\uplus^V_{({{i}k},{z}\rightarrow {i^{\ast\ast}}k^\ast,z^\ast)} =\uplus^V_{{{i}k,z}\rightarrow {i^\ast}k,z^\ast}  \times  \uplus^V_{{i^\ast}k,z^\ast \rightarrow {i^{\ast\ast}}k^\ast,z^\ast}
		\end{aligned} 
\end{equation}
Eq. (\ref{threat2}) specifies server vulnerability based unauthorised access ($\Theta^{vul}_{{ik,z} \rightarrow S_{k} }$) from $V_{ik,z}$ to server $S_{k}$ during time $\{t_1, t_2\}$, where $ \uplus^S_{ik,z}$ is evaluated using Eq. (\ref{thresholdthreat}).  The relation between $V_{i,z}$ and $k^{th}$ server ($S_{k}$) is `suspicious' if available throughput (${TP}_{V_{ik,z}}^{avl}$) and bandwidth (${BW}_{V_{ik,z}}^{avl}$) are lesser than guaranteed threshold ($\mathds{TH}_{v_{i,z}}^{gtd}$), otherwise, `non-suspicious'.
\begin{equation} \label{threat2}
\int_{t_1}^{t_2} \Theta^{vul}_{{ik,z} \rightarrow S_{k} } dt= {\int_{t_1}^{t_2} \omega_{ik,z} \times \uplus^S_{{ik,z} \rightarrow S_{k}}}dt \quad \forall \{i\} \in q, k \in p 
\end{equation}
\begin{equation}
\uplus^S_{{ik,z} \rightarrow S_{k}} = 	\begin{cases}
$1$, & {If({{{TP}}_{V_{ik,z}}^{avl} \wedge {BW}_{V_{ik,z}}^{avl}) \times \delta_t \le {\mathds{TH}_{V_{ik,z}}^{gtd}} {}}} \\
$ 0$, & { {\textit{otherwise}} } 
\end{cases} \label{thresholdthreat}
\end{equation}
The guaranteed threshold (i.e., $\mathds{TH}_{v_{i,z}}^{gtd}$) of performance parameters: throughput ($TP$) and bandwidth ($BW$) of $i^{th}$ VM of $z^{th}$ user is defined as $\mathds{TH}_{V_{ik,z}}^{gtd} =\{\mathds{TH}^{{TP}}_{V_{ik,z}} \wedge \mathds{TH}^{{BW}}_{V_{ik,z}}\}$. The total security breaches information ($\Theta^{DC}_z$) of $z^{th}$ user can be compiled by applying Eq. (\ref{threat3}):
\begin{equation}
\label{threat3}
\resizebox{0.5\textwidth}{!}{$ 
	\begin{aligned}
	\int_{t_1}^{t_2} \Theta^{DC}_{z } dt= \int_{t_1}^{t_2}\Theta^{col}_{ik,z \rightarrow {i^\ast}k,z^\ast} dt + \int_{t_1}^{t_2}\Theta^{cas}_{({i^\ast}k,z^\ast\rightarrow {i^{\ast\ast}}k^\ast,z)}dt+\int_{t_1}^{t_2}\Theta^{vul}_{ik,z} dt
	\end{aligned} $}
\end{equation}

The attack coverage ($\mathds{A}_{coverage}^{z^\ast}$) by ${z^\ast}^{th}$ malicious user is defined as the ratio of the number of malicious links ($	\mathds{L}^{Mal}_{i^\ast, z^\ast}$) between malicious VM ($ V^{Mal}_{{z^\ast}}$) and benign VMs ($V_z$) i.e., $|\sum_{k=0}^{p} \sum_{i^\ast=0}^{m^\ast} \mathds{L}^{Mal}_{i^\ast k, z^\ast}|$ and total number of malicious VMs ($Tot\_V^{z^\ast}$) during time interval \{$t_1$, $t_2$\}.  $\mathds{A}_{coverage}^{z^\ast}$ is computed using Eq. (\ref{attackcoverage1}) and $\mathds{L}^{Mal}_{i^\ast, z^\ast}$ are determined using Eq. (\ref{mal_links}).

 \begin{equation} \label{attackcoverage1}
\int_{t_1}^{t_2}	\mathds{A}_{coverage}^{z^\ast} dt = \int_{t_1}^{t_2}\frac{| \sum_{k=0}^{p} \sum_{i^\ast=0}^{m^\ast} \mathds{L}^{Mal}_{i^\ast k, z^\ast}|}{Tot\_V^{z^\ast}} dt
\end{equation}

\begin{equation}
\label{mal_links}
\resizebox{0.41\textwidth}{!}{$ 
	\begin{aligned}
\int_{t_1}^{t_2}	\mathds{L}^{Mal}_{i^\ast, z^\ast} dt= 
\int_{t_1}^{t_2}\mathds{L}^{V_{i^\ast,z^\ast}}-(\mathds{L}^{V_{i^\ast,z^\ast}} \cap \mathds{AL}^{V_{i^\ast,z^\ast}}_{legal})  dt
	\end{aligned} $}
\end{equation}
Monitor utilizes VLAM, IVCL, and Eqs. (\ref{linkthreat})-(\ref{threat3}) to identify malicious VM ($V^{Mal}$) and transmits the knowledge of $V^{Mal}$ among all the neighbouring servers to terminate all the malicious links before the propagation of the network security threat at extreme level. Further, all the malicious VMs are penalized by suspending them while notifying $U_{A}$.

\section{Congestion Avoidance and VM Allocation}
 A machine learning based predictor is deployed to analyse resource requirements of VMs for execution of future applications and estimate the bandwidth hog in real time. The future applications are grouped into $N$ clusters depending on their predicted bandwidth usage which guides the allocation of VMs to the selected network-efficient servers while mitigating the congestion proactively. Let $\mathds{T}_\sigma$ be the deviation of network traffic from the estimated traffic ($\mathds{T}$) over duration $\delta_t \in$ \{$t_1$, $t_2$\}, $\mathds{T}_\sigma^{thr}$ and $\delta_t^{thr}$ are traffic deviation and time-period thresholds respectively. Eq. (\ref{uba}) investigates real time security threat in data centre ($\Theta^{con}_{dc}$) due to network congestion.

 \begin{equation}
 \label{uba}
 \int_{t_1}^{t_2}\Theta^{con} dt= \begin{cases}
 1, & {If(\mathds{Tr}_\sigma \times \delta_t > \mathds{T}_\sigma^{thr} \times \delta_t^{thr})} \\
 -1, & {If(\mathds{Tr}_\sigma \times \delta_t < 0)} \\
 0, & {\text{otherwise}}  
 \end{cases}	
 \end{equation}
 If \textit{congestion is anticipated } (i.e., $\Theta^{con}_{dc}$=$1$), the network traffic is diverted across multiple paths and assigned to servers reserved for handling network hogs. Similarly, if \textit{underload} ($\Theta^{con}_{dc}$=$-1$) is detected, the load is shifted from underloaded to network-efficient servers; otherwise, the traffic is \textit{normal}.
 A neural network based predictor is periodically trained with historical and live resource utilization of VMs executing different applications which allows extraction of useful patterns and learning of correlations and helps to predict resources (viz. CPU, memory and bandwidth) utilization in real time. The predicted applications are grouped into $N$ clusters to filter bandwidth hogs (which may generate congestion based security attacks) and schedule them on selected network-efficient servers. K-means clustering is applied to keep the clusters as different as possible while grouping predicted applications of similar bandwidth usage so that the sum of the squared distance ($\mathds{G}$) between their bandwidth requirement and centroid of the cluster is minimum by applying Eq. (\ref{eq.cluster}), where ${w}_{ic}$ defines mapping between bandwidth of $i^{th}$ application ($BW_i$) and centroid ($\mu_c$) of $c^{th}$ cluster.
\begin{equation}
\label{eq.cluster}
\mathds{G}=\sum_{i=1}^{m}\sum_{c=1}^{N}{{w}_{ic}{|{BW}_i- \mu_c|}^2}
\end{equation}
 The deployment of $i^{th}$ VM ($V_i$) on $j^{th}$ server ($S_j$) must satisfy the resource (CPU ($C$), Memory ($M$), Bandwidth (${BW}$)) capacity constraints stated in Eq. (\ref{VMP}), where $V_i^{\mathds{R}}$ specifies resource requirement of $V_i$ and $S_k^{\mathds{R}}$ is available resource capacity of server $S_k$. 
\begin{equation}
\sum_{i=1}^{q}{V_i^{\mathds{R}} \times \omega_{ik} \le S_k^{\mathds{R}}}; \quad \forall_k \in \{1, p\}, {\mathds{R}} \in \{C, M, BW\} \label{VMP}
\end{equation}

The resource utilization ($RU$) of data centre is evaluated by using Eqs. (\ref{RU1}) and (\ref{RU2}), where $\mathds{N}$ is the number of monitored resources, $RU^C$, $RU^M$ and $RU^{BW}$ are CPU, memory and bandwidth utilization of server. If $k^{th}$ server $S_k$ is active i.e., it hosts VM ($\eta_k$ = 1), otherwise, it is inactive ($\eta_k$ = 0).  

\begin{equation}
RU^{DC}= \frac{\sum_{k=1}^{p}{RU_k^{C}} + \sum_{k=1}^{p}{RU_k^{M}} + \sum_{k=1}^{p}{RU_k^{BW}}}{|\mathds{N}| \times \sum_{k=1}^{p}{\eta_k} } \label{RU1}
\end{equation}

\begin{equation}
RU_k^{\mathds{R}} = \frac{\sum_{i=1}^{q}{\omega_{ik}} \times V_i^{\mathds{R}}}{S_k^{\mathds{R}}} \quad \forall_k \in \{1, p\}, \mathds{R} \in \{C, M, BW\} \label{RU2}	
\end{equation}
The total power consumption ($PW^{DC}$) during time-interval \{$t_1$, $t_2$\} is computed by applying Eq. (\ref{power2}), where ${PW_i}^{max}$, ${PW_i}^{min}$, and ${PW_i}^{idle}$ are maximum, minimum, and idle state power consumption, respectively of $i^{th}$ server.
\begin{equation}
PW^{DC} = 
\sum_{i=1}^{P} {[{PW_i}^{max} - {PW_i}^{min}]\times{RU} + {PW_i}^{idle}}
\label{power2}
\end{equation} 
\section{ Operational Design and Illustration}
OSC-MC utilizes historical and current bandwidth usage of different applications to anticipate the demand of bandwidth for the VM executing the application in real-time, and accordingly, balances the load on different servers to mitigate any probability of network congestion based security attacks. A module named \textit{Monitor} keeps track of inter-VM links among VMs on a server in a matrix and analyses them to determine probability of co-residency and network cascading attacks.  
\subsection{Complexity Computation}
Algorithm \ref{algo-osecc} presents the detailed operational summary of OSC-MC, where the complexity of steps 1 and 2 are $O(1)$ and $O(mpq)$, respectively. VMs are allocated using \textit{First-Fit Decreasing} algorithm based on the decreasing order of predicted bandwidth so as to reduce bandwidth wastage. The steps 3-15 repeat for $t$ intervals. Steps 4-6 execute $q$ times to predict resources requirement of $q$ VMs. Step 7 calls K-Means clustering with time complexity of $O(m^2)$. Again, the step 8 maps $q$ VMs to $p$ servers consumes complexity of $O(pq)$. Steps 9-14 execute for $q^2$ times. Hence, the total complexity comes out to be $O(m^2q^2pt)$.

\begin{figure}[!htbp]
	\removelatexerror
	\begin{algorithm}[H]
		\caption{OSC-MC: Operational Summary}
		\label{algo-osecc}
		Initialize: $List_{\mathds{U}}$, $List_{\mathds{V}}$, $List_{\mathds{S}}$\; 		
		Allocate $V_1$, $V_2$, ..., $V_q$ to $S_1$, $S_2$, ..., $S_p$ by defining a mapping $\mathds{U} \times \mathds{V}\mapsto \mathds{S}$ and generate IVCL \;		
		\For {each time-interval $\{t_1, t_2\}$}{ 
			\For {each VM $V_i$}{
			[${V_{t_2}}^{Pred(C_i, M_i, BW_i)}$] $\Leftarrow$ Resource prediction(${V}_{t_1}$) $\forall i \in [1, q]$ \;
		}
	Clustering ($C_1$, $C_2$, ..., $C_N$) $\leftarrow$ K-Means($\mathds{V}^{BW}_{t_2}$)\;
	Schedule $V_1$, $V_2$, ..., $V_q$ on servers and generate VLAM for each server\;
	\eIf{$\mathds{AL} \nsubseteq \mathds{AL}_{legal}$}{
		Identify and terminate malicious VM ($V^{Mal}$) by comparing IVCL and VLAM\;
	}
		 {Keep ${V}_i$ at same server until user terminates it\;}
		
		}
	\end{algorithm}
\end{figure}
\subsection{Illustration}
Consider four sets of inter-dependent VMs such that \{$V_1$, $V_2$, $V_{3}$, $V_{4}$\}, \{$V_{5}$, $V_6$, $V_{7}$ \}, \{$V_8$, $V_{9}$, $V_{10}$, $V_{11}$\}, \{$V_{12}$, $V_{13}$, $V_{14}$, $V_{15}$\} belong to $U_1$, $U_2$, $U_3$, and $U_4$, respectively. Taking the assumption that \{$V_{1}$, $V_{3}$, $V_{11}$\} are hosted on server $S_1$, \{$V_{2}$, $V_{7}$, $V_{12}$\} are deployed on $S_2$, \{$V_{5}$, $V_{6}$, $V_{8}$\} on $S_3$, \{$V_{4}$, $V_{9}$, $V_{14}$\} on $S_4$, and \{$V_{13}$, $V_{10}$, $V_{15}$\} on $S_5$. Let a set of VMs \{$V_8$, $V_{9}$, $V_{10}$, $V_{11}$\} of user $U_3$ be malicious. The inter-dependent VMs on same or different servers (viz. $V_{1}$ and $V_{3}$ hosted on $S_{1}$, $V_{5}$ and $V_{6}$ hosted on $S_{3}$ etc.) exchange data among them which is sensed by the malicious VMs $V_8$, $V_{9}$, $V_{10}$, $V_{11}$ to access sensitive data by establishing unauthorised links, side-channels, exploiting hypervisor vulnerability or capturing traces of valuable data from cache of server hosting them. Based on given allocation, colocation attacks occur on $S_1$, $S_3$ and $S_5$ (i.e., $\Theta^{col}=60\%$) and cascading attack can propagate via multiple connections (i.e., $\Theta^{cas}=100\%$). To eliminate such attacks, OSC-MC employs an exclusive \textit{Monitor} on each server to surveil all the current inter-VM links established on a server in VLAM, compare them with the authorised link information in IVCL and terminate all malicious links and suspend malicious VMs $V_8$, $V_{9}$, $V_{10}$, $V_{11}$.

\section{Performance Evaluation and Discussion}
The simulation experiments are executed on a server machine assembled with two Intel\textsuperscript{\textregistered} Xeon\textsuperscript{\textregistered} Silver 4114 CPU with 40 core processor and 2.20GHz clock speed, deployed with 64-bit Ubuntu 16.04 LTS having main memory of 128 GB in Python 3.1. The data centre environment is set up with quad and dual core processor types of server having CPU: 2000 MIPS, RAM: 2048 MB, BW: 10,000. Two types of VMs having (500, 512, 1000) and (1000, 1024, 1000) as CPU, RAM and BW utilization respectively are used. Three-layered \textit{feed-forward neural network} is used for resource prediction and the \textit{hypervisor vulnerability score} associated to each server is generated in the range [0, 10] randomly where a value $\ge7$ indicates high probability of attack. We experimented with benchmark \textit{Bitbrains} dataset containing resources (CPU, Memory and Bandwidth) usage of more than 1400 VMs with a periodic interval of 5 minutes, the number of users  equals to one-third of total number of VMs, and inter-VM links are randomly generated during run-time.
\par 
OSC-MC is compared to \textit{Security Embedded Dynamic Resource Allocation} (SEDRA) \cite{sedra}, \textit{Previously Selected Server First} (PSSF) \cite{pssf} and \textit{Secure and Energy-Aware Load Balancing} (SEA-LB) \cite{sealb}. PSSF maintains list of cloud users and allocates their requested new VMs on the same server which have previously hosted their VMs. SEDRA considers performance of network traffic and inter-VM links to detect and mitigate VM threats by utilizing a random tree classifier. SEA-LB assigns VMs to energy-efficient servers subject to maximum resource utilization, minimum power consumption and side-channel attacks by using a modified multi-objective genetic algorithm approach. The number of shared servers among users are minimized to provide security.
\subsection{Results}
Table \ref{table:self} shows achieved results of different performance metrics for OSC-MC with various size of data centre and varying number of benign and malicious users. The average number of malicious links ($\mathds{L}^{Mal}$), unauthorized inter-VM access ($\Theta$), and VMs (${V}^{Mal}$) increases unevenly with the number of malicious users and growing size of the data centre. Also, the average number of $\mathds{NW}^{Hogs}$ varies with dynamic network traffic and size of the data centre. The mean of $RU$ (\%) ranges from 61\% to 64\% while $PW$ ($KW$) scales up with rising size of the data centre.
 
\begin{table}[!htbp]
	\centering
	
	\caption[Table caption text] {Performance metrics for OSC-MC}
	\label{table:self}
	
	\resizebox{9cm}{!}{
		\begin{tabular}{|l|c|c|c|c|c|c|c|}
			\hline
			
		VMs&$U$\# ($U_{A}$\%)&$\mathds{L}^{Mal}$& $V^{Mal}$&$\Theta$ (\#)&$\mathds{NW}^{Hogs}$&$RU$ (\%)& $PW$($KW$)\\
			\hline \hline
			
			\multirow{3}{*}{200}&66 (5)&43 &6&43 &13&63.45 &9.58 \\ \cline{2-8}
			&66 (20)&58&17&58&12&62.06 &9.96\\ \cline{2-8}
			&66 (90)&306&131&306&16& 62.48&9.29\\ 	\hline \hline
			
				\multirow{3}{*}{500}&166 (5)&12 &9&12 &22&61.73 &22.73\\ \cline{2-8}
			&166 (20)&109&64 & 109&23&60.42 &22.58\\ \cline{2-8}
			&166 (90)&1053&319&1053&21& 61.48&23.02\\ 	\hline \hline
				\multirow{3}{*}{800}&266 (5)&77 &34&77 &42&62.23&44.13 \\ \cline{2-8}
			&266 (20)&456& 116&456 &39&62.41 &43.47\\ \cline{2-8}
			&266 (90)&1297&518&1297&41& 62.07&43.18\\ 	\hline
			\hline
			
			\multirow{3}{*}{1100}&366 (5)&99 &43&99 &46&63.20 &78.34 \\ \cline{2-8}
			&366 (20)&387& 161&387&41&64.05 &78.47\\ \cline{2-8}
			&366 (90)&1918&706&1918&47& 63.88&78.18\\ 	\hline
	\end{tabular}}
	
\end{table}

\subsection{Comparison}
\begin{figure}[!htbp]
 	\centering
 	\includegraphics[width=0.7\linewidth]{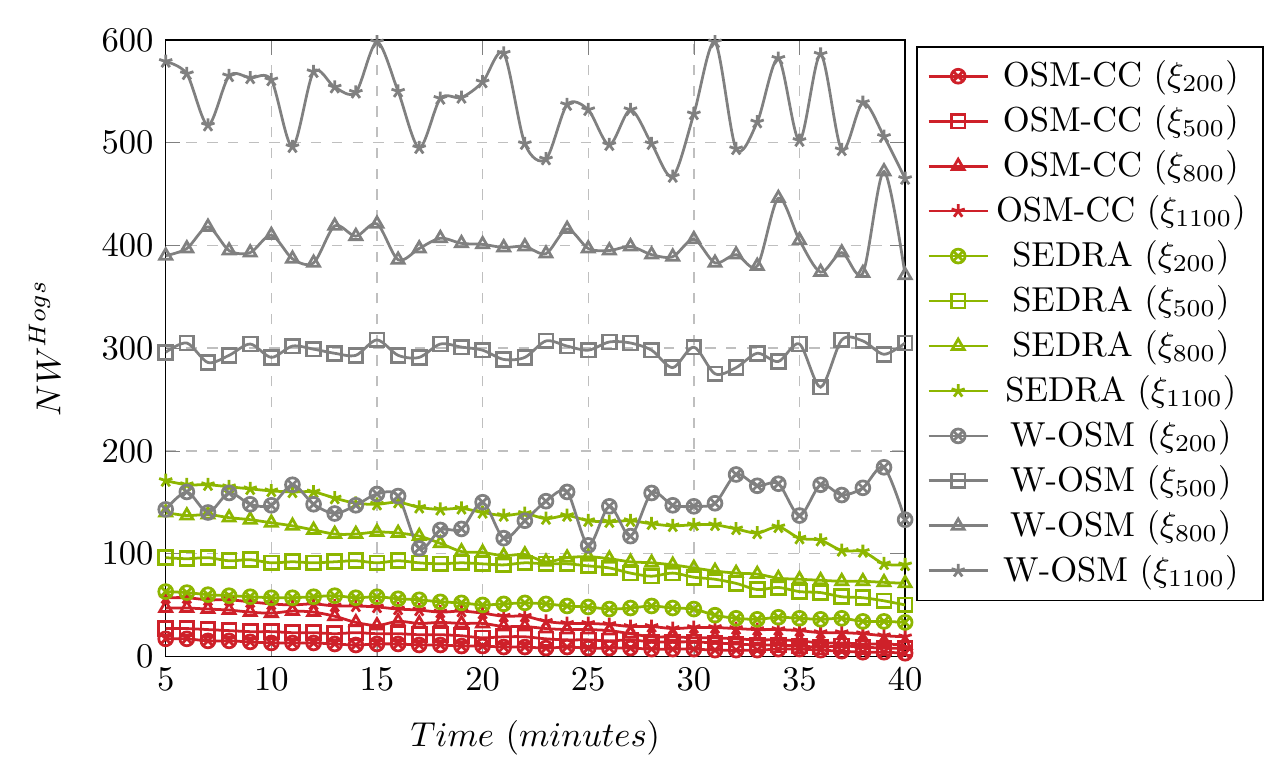}
 	\caption{Network hogs}
 	\label{fig:networkhogs}
 \end{figure}

\begin{figure}[!htbp]
	\centering
	\includegraphics[width=0.7\linewidth]{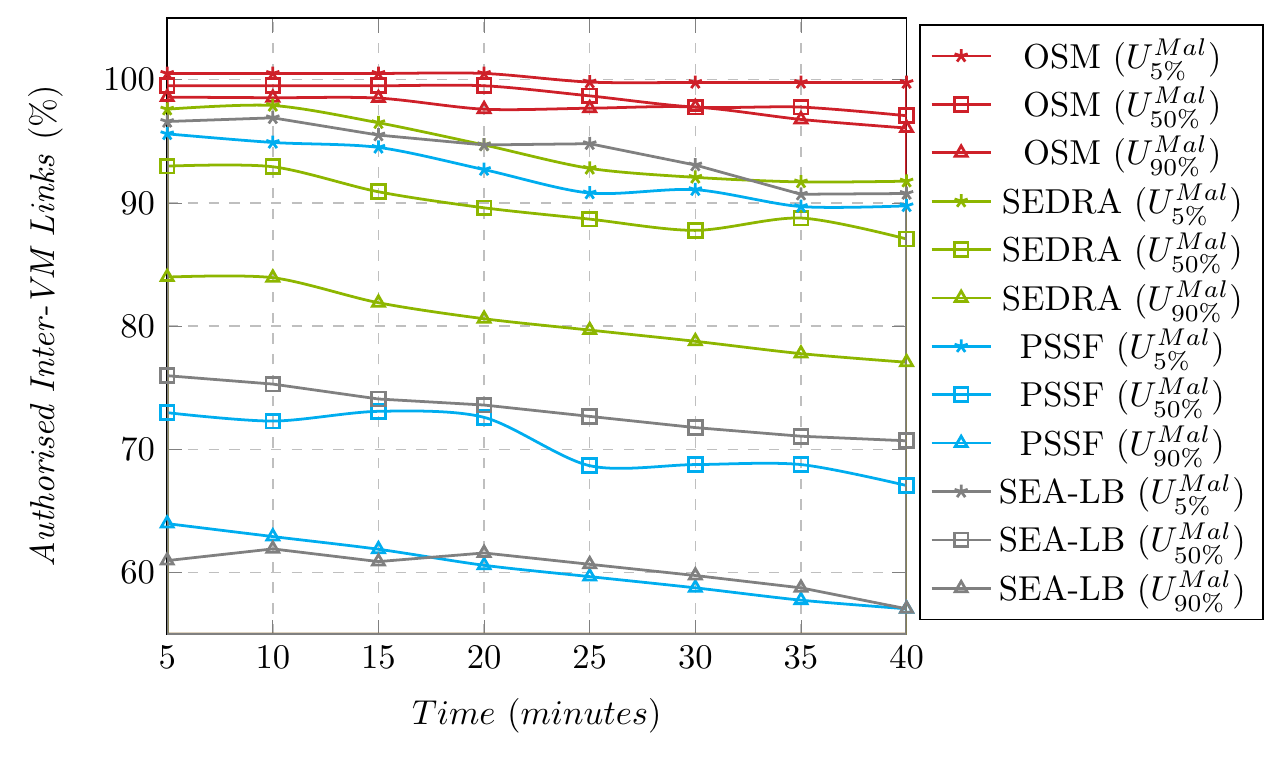}
	\caption{Authorised Inter-VM Links over time}
	\label{fig:mal_links}
\end{figure}

The number of network hogs obtained with OSC-MC, SEDRA and without-OSC-MC (W-OSC) for data centre of sizes 200 VMs ($\xi_{200}$), 500 VMs ($\xi_{500}$), 800 VMs ($\xi_{800}$), and 1100 VMs ($\xi_{1100}$) are shown in Fig. \ref{fig:networkhogs}. The number of hogs in case of OSC-MC, is below 70 for every size of data centre which scales down with the time period because of periodic learning of neural-network predictor in real-time. OSC-MC reduces congestion up to 66.46\% and 89.94\% over SEDRA and W-OSC, respectively for $\xi_{1100}$.
 Fig. \ref{fig:mal_links} compares inter-VM communication links over period \{$t_1$, $t_2$\} observed for OSC-MC with that of SEDRA, SEA-LB, and PSSF for varying number of $U_{A}$. The authorised link percentage $\mathds{AL}$ (\%) has reached up to $100\%$ in case of OSC-MC while it is observed to be 77\%-97\%, 57\%-97\%, and 57\%-96\% for SEDRA, SEA-LB, and PSSF, respectively. 

\begin{figure}[!htbp]
	\centering
	\includegraphics[width=0.7\linewidth]{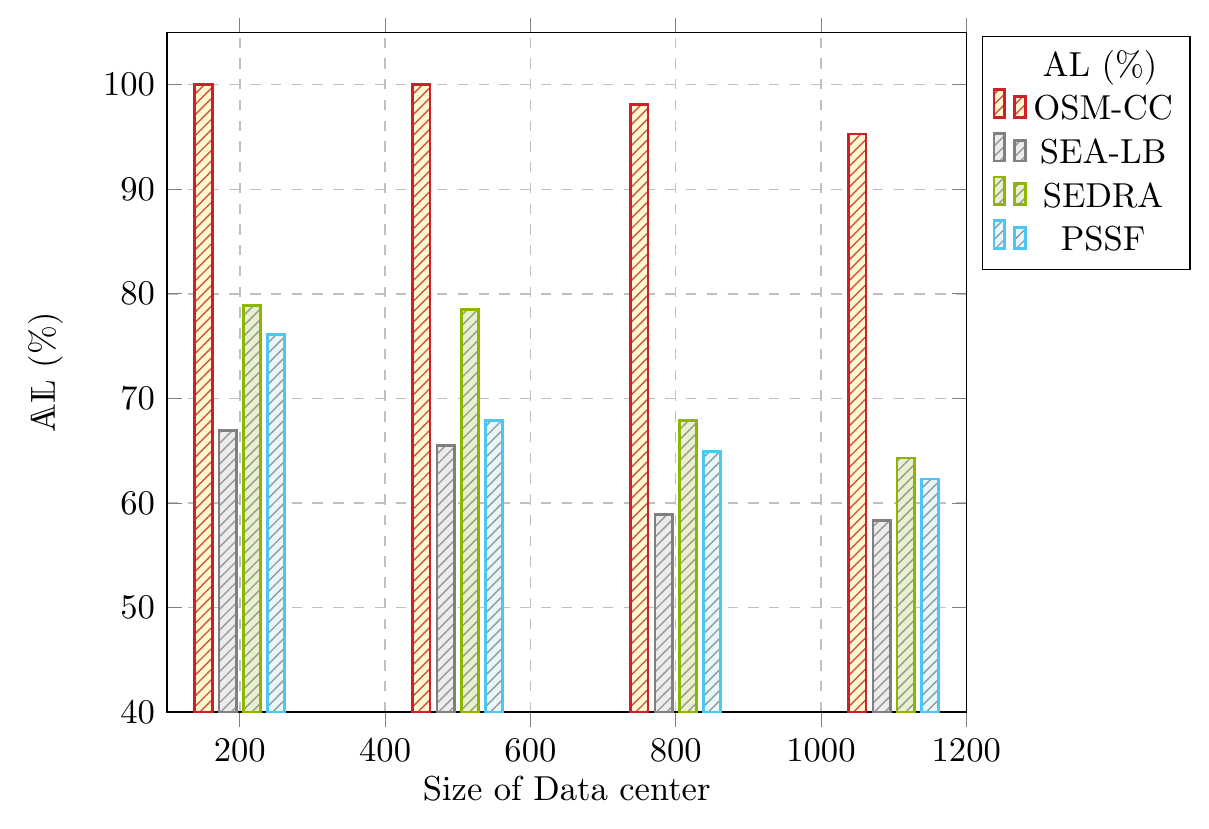}
	\caption{Average authorised access percent for different size of data centre  }
	\label{fig:securitybreachru}
\end{figure}

Fig. \ref{fig:securitybreachru} compares $\mathds{AL}$ ($\%$) with respect to various sizes of data centre where, OSC-MC detects and mitigates almost every $\mathds{L}^{Mal}$ and $V^{Mal}$ before the completion of enough time required for occurrence of actual security threat. OSC-MC scales down active $V^{Mal}$ and $\mathds{L}^{Mal}$ among VMs by monitoring mutual inter-relations among them and raises the average legal access percentage by 32.1\%, 34.5\% and 21.5\% against PSSF, SEA-LB, and SEDRA, respectively for 500 VMs. 
 \par Table \ref{table:comparison} compares power consumption ($PW$) and resource utilization ($RU$) by OSC-MC with other three schemes; where $PW$ is reduced up to 39.31\%, 31.95\% and 17.65\% against PSSF \cite{pssf}, SEA-LB \cite{sealb}, and SEDRA \cite{sedra}, respectively for 1100 VMs. $RU$ is evaluated using Eqs. (\ref{RU1}) and (\ref{RU2}) which increases as: \textit{PSSF} $<$
 \textit{SEDRA} $<$ \textit{SEA-LB} $<$\textit{OSC-MC}. $RU$ is independent of the size of data centre and varies in the constant ranges of 55\%-56\%, 59\%-60\%, and 61.5\%-62.5\% for PSSF, SEDRA, and OSC-MC, respectively. The reason behind above shown results is the online prediction of required resources beforehand that assisted in consolidation of VMs on least number of active servers.

\begin{table}[!htbp]
	\centering	
	\caption[Table caption text] {Power consumption and Resource utilization}
	\label{table:comparison}
	
	\resizebox{7cm}{!}{
		\begin{tabular}{|l|c|c|c|c|c|}
			\hline
			
			VMs&Metrics &SEA-LB& PSSF&SEDRA&OSC-MC\\
			\hline \hline
			
			\multirow{2}{*}{200}&$PW$ ($KW$)&30.92 &32.13&14.23 &9.17  \\ \cline{2-6}
			&$RU $ (\%)&60.12&58.12&59.12&62.06\\ 	\hline \hline 
		\multirow{2}{*}{500}&$PW$ ($KW$)&65.66 &74.29&51.24 &22.02  \\ \cline{2-6}
		&$RU $ (\%)&60.92&58.62&59.92&61.73\\ 	\hline \hline 	
			
			\multirow{2}{*}{800}&$PW$ ($KW$)&84.36 &93.39&69.09 &43.35 \\ \cline{2-6}
		&$RU $ (\%)&60.81&58.94&59.94&62.06\\ 	\hline \hline 
		\multirow{2}{*}{1100}&$PW$ ($KW$)&129.67 &143.58&107.28 &88.35 \\ \cline{2-6}
		&$RU $ (\%)&60.36&58.31&59.46&61.28\\ 	\hline 	
	\end{tabular}}
	
\end{table}

\section{Conclusion}
A novel online security ingrained cloud communications model: OSC-MC is proposed that monitors inter-VM relations and detects malicious VMs to mitigate security breaches. OSC-MC incorporates machine learning based congestion prediction and maintains IVCL and VLAM to determine and terminate suspicious inter-VM links. It avoids security breaches due to known and unknown VMs and substantially decreases malicious congestion, network cascading threats, and co-residency threats. The achieved results support influential performance of OSC-MC against the compared approaches.


%



\section*{Acknowledgment}
This work is financially supported by National Institute of Technology, Kurukshetra, India.

\end{document}